# A NEW CONCEPT OF TECHNOLOGY WITH SYSTEMIC-PURPOSEFUL PERPSECTIVE: THEORY, EXAMPLES AND EMPIRICAL APPLICATION

*Mario Coccia*

1) CNR -- NATIONAL RESEARCH COUNCIL OF ITALY
Collegio Carlo Alberto, Via Real Collegio, 30-10024 Moncalieri (Torino, Italy)

2) YALE UNIVERSITY, 310 Cedar Street, Lauder Hall, New Haven, CT 06510, USA

*E*-mail: mario.coccia@cnr.it

**Abstract.**

Although definitions of technology exist to explain the patterns of technological innovations, there is no general definition that explain the role of technology for humans and other animal species in environment. The goal of this study is to suggest a new concept of technology with a systemic-purposeful perspective for technology analysis. *Technology here is a complex system of artifact, made and/or used by living systems, that is composed of more than one entity or sub-system and a relationship that holds between each entity and at least one other entity in the system, selected considering practical, technical and/or economic characteristics to satisfy needs, achieve goals and/or solve problems of users for purposes of adaptation and/or survival in environment.* Technology changes current modes of cognition and action to enable makers and/or users to take advantage of important opportunities or to cope with consequential environmental threats. Technology T, as a complex system, is formed by different elements given by incremental and radical innovations. Technological change generates the progress from a system T1 to T2, T3 …. driven by changes of technological trajectories and technological paradigms. Several examples illustrate here these concepts and a simple model with a preliminary empirical analysis shows how to operationalize the suggested definition of technology. Overall, then, the role of adaptation (i.e. reproductive advantage) can be explained as a main driver of technology use for adopters to take advantage of important opportunities or to cope with environmental threats. This study begins the process of clarifying and generalizing, as far as possible, the concept of technology with a new perspective that it can lay a foundation for the development of more sophisticated concepts and theories to explain technological and economic change in environment.

# INTRODUCTION AND GOAL OF THIS STUDY

This study has two goals. The first is to define the technology with a systemic-purposeful perspective and suggest properties of its behavior. The second is to operationalize this concept to show practical applications for management of technology and economics of innovation.

The concept of technology plays an important role in the economic and social change of human societies (Basalla, 1988; Berg et al., 2019; Coccia, 2019, 2019a; Freeman and Soete, 1987; Hosler, 1994; Moehrle and Caferoglu, 2019; Nelson and Winter, 1982). Hickman (2001) claims that technology is a central feature of the human-nature, and human-human. Hickman (2001, pp. 40-41) suggests that technology is a set of techniques, in particular as inquiry into techniques, tools, and artifacts in which techniques are habitual and traditional ways of dealing with things. According to Hickman (2001, p. 183), technology can be understood as: "the intelligent production of new tools, including conceptual and ideational ones, for dealing with problematic situations". In particular, the study by Hickman (2001) differentiates among tools, techniques and artifact, as follows:

> Techniques, tools, and artifacts in fact make up a kind of ascending series of more or less stable "spaces" within which human beings make-that is, produce-their world. But I am not sure that we should call an inquiry into them, or the processes by which and within which they arise, technology. The critical point here is that each space is, or relies upon, or is constituted by embodied knowledge (as quoted by Innis, 2003, p. 35, original emphasis).

Moreover, Hickman (2001, p. 98), considering the theory of inquiry by Dewey (1938, 1958), states that: "Progress is rather a cycle of production: this includes the production of new significances, the production of new feelings, the production of new means of enjoying, and the production of new techniques of production" (cf., Pacey, 1999).

In economics, patterns of technology emerge and evolve with technological paradigms and trajectories in specific economic, institutional and social environments (Dosi, 1988). Hosler (1994, p. 3, original italics) argues that technology and its development is, at least to some extent, influenced by "technical *choices*", which express social and political factors, and "technical *requirements*", imposed by material properties. Sahal (1981) argues that technology has manifold dimensions, ranging from an object of material culture to an organized group of applied scientific knowledge.

Brey (2009) argues that general public knows what technology is and how it can support human activity. However, the concept of technology remains ambiguous and ill defined. The main goal of this paper is to suggest a theoretically and analytically comprehensive definition of technology. The approach of the study here is based on a systemic-purposeful perspective that may explain and generalize, whenever possible, aspects of technology in human societies and environment. The theoretical and empirical analyses here hint at general properties of technology to clarify its origins and how it continues to evolve in socio-ecological environments. This new theoretical framework lays a foundation for the development of more sophisticated concepts and theories that explain technological coevolution, technical and economic change in human society and environment.

**FROM ARTIFACTS, INSTRUMENTS TO TECHNOLOGY AS MEANS OF ADAPTATION OF HUMANS AND OTHER ANIMAL SPECIES IN ENVIRONMENT**

Biro et al. (2013) argue that tool use is a component of human behavior. The benefits of tool are self-evident and given by extending control over our environment, by increasing energetic returns and by buffering ourselves from potentially harmful influences. The dependence of people on things that they make and use unifies all mankind, such that material objects are essential for human life. When human condition emerges, our predecessors are makers of tools and this activity has led to the origins of different technology (cf. also, Tria et al., 2014; Sahal, 1981).

Oswalt (1976, p. 18) explains the origin of technology, with an anthropological analysis that differentiates between naturefacts, artifacts and instruments. *Naturefact* is based on natural forms that are used in place or withdrawn from a habitat, without prior modification by creatures. Naturefacts are the logical basis from which all man-made productions may have originated, such as hand weapons. The term creatures, in the definition of naturefact, suffices to isolate the users and is introduced to accommodate any animals employing a natural object. Examples of naturefact are the stone as missile for birds, stick to dig roots, etc. (Oswalt, 1976, p.21ff; McGrew, 2013).

The *artefact* or *artifact* in American usage is a simple object (e.g., a tool) showing human workmanship or modification as distinguished from a natural object. Fragaszy et al. (2013) highlight how artefacts create rich learning opportunities for young individuals. Examples of artifacts used by aboriginal are thorn for septum pierces, leaf for body cleaner, etc. (Oswalt, 1976, p. 26). Clarke (1968, p. 186) defines artifact as any object modified by a set of humanly imposed attributes, whereas Titiev (1963, p. 632) considers artifact "any object that is consciously manufactured for human use". Oswalt (1976, p. 24) suggests a comprehensive definition: "an artifact is the end product resulting from the modification of a physical man in order to fulfil a useful purpose". This definition is general because both human and other animal species (e.g., Caledonian crows) can make things to be used in food-getting situations (cf., St Amant and Hortonm, 2008; Tolman, 1932). In fact, people are not the only makers of artifacts. Birds fashion nets and beavers build dams are acceptable within the scope of the suggested definition of artefact.

McGrew (2013) argues that the chimpanzee is well-known in both nature and captivity as an impressive maker and user of tools, but recently the New Caledonian crow has been championed as being equivalent or superior to the ape in this elementary technology. In particular, McGrew (2013) performs a direct comparison between New Caledonian crows and chimpanzees, the two non-human species typically considered the most 'advanced' animal tool users. Along some axes of comparison tool use, New Caledonian crows' approaches surpass chimpanzee technology (e.g., manufacture of hooked foraging tools), in others the apes register higher counts of observed behaviors. In general, naturefacts and artifacts are composed of materials and have a physical form. The naturefact-artifact distinction is made to clarify the ways in which natural forms were used.

The word *instrument* identifies hand-manipulated subsistants that customarily are used to impinge on masses incapable of significant motion and relatively harmless to people (Oswalt, 1976). Examples of instruments are digging stick, ax for procuring animals, etc. to obtain plant and animal products as food (Oswalt, 1976, p. 70). Instruments can be extensions of human hands and/or competitors with hands. Moreover, the evolution of material culture is based on application of instrument technology used for the cultivation of plants as food that led to surpluses and remarkable elaborations in other aspects of human life.

Biro et al. (2013) argue that the performance of skilled tool users, it provides further important clues to the potential lifetime adaptive benefits of behavior. For instance, Haslam (2013) states that among the great apes, individuals in captivity exhibit a greater range of tool-related behaviors than their counterparts in the wild. Haslam (2013) also suggests a number of environmental and social factors that could account for this effect, such as increased free time and increased access to both materials and individuals, which are skilled in using them as tools.

Collard et al. (2013) find evidence for the "(environmental) risk hypothesis" that the use of more specialized and elaborate tools may buffer against the risks of resource failure, leading to richer tool kits in riskier environments. In general, the interaction of physical and social environmental variables drives technological evolution, suggesting that these variables should not be considered in isolation (cf. Coccia, 2018a; Haslam, 2013; Kline and Boyd, 2010; Henrich, 2004). Biro et al. (2013) argue that trajectories of tool-use development show immense variation across species: some appear as genetically fixed action patterns, some are acquired through individual learning and some are cases of social learning. In particular, for both individually and socially acquired behaviors (analyzed by Humle et al., 2009), the physical and/or the social environment must present sufficient opportunities—or sufficient necessity (see Haslam, 2013; Collard et al., 2013, 2011)—to promote individuals' tool-use learning, notwithstanding any possible morphological or cognitive prerequisites. Teschke et al. (2013, 2011) analyze the role of cognition either as a domain-general pre-adaptation to flexible tool use or as a more domain specific adaptation that has evolved to support increasingly sophisticated forms of tool use. Comparative studies examine whether naturally tool-using species possess cognitive capabilities that differ from those of their close, naturally non-tool using relatives. Some studies compare physical-cognition and general learning tasks presented to both tool-using New Caledonian crows and non-tool-using carrion crows. Teschke et al. (2011) reveal that the tool-using species 'outperforms' its non-tool using counterpart on tasks involving physical cognition (but not on those testing general-learning abilities). However, results should be treated cautiously.

In general, humans are by far the most versatile tool users in existence. Marzke (2013) and Hashimoto et al. (2013) reveal long-term effects of tool technology on human biology. The advent of stone-tool use was undoubtedly a key event in our own lineage's evolution, eventually leading to the establishment of humans as the most successful tool users on the planet. The analysis by Marzke (2013) shows that the evolution of human hand induces features for grip and stress-accommodation that are necessary to support stone-tool manufacture. Iriki et al. (2001) and Maravita et al. (2002) have provided evidence that with tool-using tasks, the brains of both humans and monkeys perceive tools as extensions of the individuals' bodies to solve problems.

Hence, tools represent the direct between animal and its environment and they play a vital role for adaptation in environment. Elongated tools are found both within the hominin line and among non-human animals (including the types of stick tools manufactured and used by chimpanzees and New Caledonian crows). Gowlett (2013) argues that elongation represented one end of a continuum of shapes that serve specific needs in different tasks.

Overall, then, the role of adaptation (i.e. reproductive advantage) can be as an ultimate explanation for tool, artifact, instrument and technology use to take advantage of important opportunities or to cope with environmental threats.

**CRITIQUE OF CURRENT CONCEPTS OF TECHNOLOGY**

People know technology and can discern natural things from human-made ones. Technology can either be natural or be human-made, i.e., unnatural (Biro et al., 2013; Nelson 1932). Volti (2009) argues that the word "Techne" is widely accepted to mean "skill" and "art" (cf., Skrbina, 2015). The usage of words incorporating this root implies that a certain amount of skillfulness or artistry must be involved in that to which they refer. Volti (2009, p. 6) defines technology as "a system created by humans that uses knowledge and organization to produce objects and techniques for the attainment of specific goals". Examples are laser, television, computer, etc. In short, technology is a system that allows to produce objects and perform techniques to achieve goals (Carroll, 2017).

Bigelow (1829) states that technology is "understood to consist of principles, processes, and nomenclature of the more conspicuous arts, particularly those which involve applications of science, and which may be considered useful, by promoting the benefit of society, together with the emolument of those who pursue them". Arthur (2009, pp. 18-19) argues that: "Technologies somehow must come into being as fresh combinations of what already exists." This combination of components and assemblies is organized into systems or modules to some human purpose and has a hierarchical and recursive structure: i.e., "technologies … consist of component building blocks that are also technologies, and these consist of subparts that are also technologies, in a repeating (or recurring) pattern" (Arthur, 2009, p. 38). In addition, Arthur (2009) claims that technological evolution is based on "supply" of new technologies assembling existing components and on "demand for means to fulfill purposes, the need for novel technologies." (cf., Wagner, 2011; Wagner and Rosen, 2014; Ziman, 2000). Other scholars suggest that advances of technology are driven by solving consequential problems during the engineering process (Coccia, 2017; cf., Dosi, 1988; Usher, 1954) and by goals of purposeful organizations in specific socioeconomic contexts (Coccia, 2017a).

The concept of technology has a vast literature that can be categorized in three groups. *Firstly*, the economic conception of production function, *secondly* the Pythagorean concept of technology based on patents statistics and chronologies of innovations and *finally* the systems concept of technology conceived in terms of technical performance of its characteristics. However, these different viewpoints have a lot of limitations.

☐ *Neoclassical specification of economic concept of technology*

Firms produce outputs from various combination of inputs. The set of all production plans is the set of production possibilities of firms and denoted by *Y* that provides a complete description of the technological possibilities facing the firm. The description of production sets is to list the possible production plans. Varian (1984) shows the example of the production of an input using two inputs 1 and 2. This production can be done with two different techniques:

*Technique A*: 1 unit of good 1 and 2 units of good 2, it produces 1 unit of output.

*Technique B*: 2 units of good 1 and 1 units of good 2, it produces 1 unit of output.

These engineering data are available technology. In general, many possible ways can produce a given level of outputs and can fit a curve through the possible production points. A convenient way to represent technology, in a neoclassical perspective, is by a parametric function involving unknown parameters, such as the function of Cobb-Douglas technology that any input, such as capital=*K* and labor=*L*, which satisfies the condition that $K^a L^b \geq 1$ producing at least 1 unit of output (*a* and *b* are parameters). These parametric representations of technology are convenient to analyze the production choices of firms, using calculus and algebra. This production function may be illustrated by a smooth, convex isoquant representing different techniques in the production of same output. The development of new techniques generates a shift towards the origin of isoquant. As a consequence, technological advances make possible the production of the same amount of production by a lesser amount of factors, such as capital and labor. In particular, technology, in the presence of two factors of production, such as capital and labor, can generate labor saving if it increases, and capital saving if it decreases, the capital-labor ratio in the production of a given volume of output.

However, a production function based on empirical data entails numerous difficulties (Salter, 1969). In particular, any production process involves a number of variables besides factor substitution and technical progress. Moreover, measurement of variables in the production function also generates difficulties because of heterogeneous inputs and outputs. Economists have suggested alternative approaches to the neoclassical conception of the production function, such as Kaldor (1957) proposes a technical progress function in which the growth of capital per man is associated with the growth of output per men at a given rate of change, rather than a given level of technical knowledge. However, both technical progress function and production function fail to isolate economic factors from technical ones. Another limitation is that these approaches lack of a concept of technology *per se.*

❑ *The Pythagorean concept of technology*

The Pythagorean concept of technology is based on approaches of the history of science, sociology, biology, etc. (Schmookler, 1966). The concept of technology and technology evolution is a count of relevant events, such as number of patents (Jaffe and Trajtenberg, 2002). The technical activity is measured with patent statistics and the chronology of major and minor inventions. The advantages are that data of patents are available. However, patented inventions do not provide information if a new device is suitable for production and commercialization. Moreover, patents do not consider the phase of development of technology and that many inventions are not patented for various reasons, such as the inadequacy of patent protection, legal problems, etc. The alternative approach is the chronology of innovations, which assigns dates of occurrence of major and minor innovations (cf. Sahal, 1981). However, this alternative approach lacks of a theoretical framework that distinguishes different types of innovations, such as incremental and radical innovations, etc. Finally, this approach shows its relevance on the origin rather than the development of technology.

❑ *The systems concept of technology*

This approach focuses on functional characteristics of technology. The measurement of technological advances is due to change of variables based on technical function, such as fuel-consumption efficiency of a device (Sahal, 1981). This approach was applied to analyze the advances and capabilities of military technology (Alexander and Nelson, 1973; Martino, 1985; Knight, 1985; Koh and Magee, 2006). The advantages of systems approach to technology are that functional measures of technology are clearly defined and objectively measured, such as the thermal efficiency of an electric power plant, fuel-consumption efficiency in horsepower per hour per gallon in farm tractor, etc. (Sahal, 1981). Moreover, functional measures of technology provide practical value for engineering and managerial decisions to increase the efficiency of a technology and as a consequence of firms. The systems approach also evaluates major and minor innovations. For instance, in the case of farm tractor the measure of fuel-consumption efficiency can show major innovations, such as the use of pneumatic tires, quality of fuels, and minor innovations, such as durable valve, piston, etc. This approach can support management of technology as well as R&D management of firms in competitive markets.

However, the systems concept of technology has also some limitations. Data of functional characteristics of technology can be difficult to gather in the presence of a multiplicity of technological advances, such as for smartphones (Coccia, 2019a). Another limitation is that this approach is better for micro analyses rather than macro ones.

Overall, then, these three approaches have been criticized because they do not clarify the understanding of all characteristics of technologies, such as drivers, evolution, purpose, adopters, etc. Therefore, current definitions of technology are not sufficiently comprehensive of this vital concept. Moreover, many current approaches neglect to acknowledge, or underemphasize the fact that both the making and use of tools does occur in animal species other than humans (Boesch and Boesch 1984; Biro et al., 2013). Overall, current definitions do not provide sufficient explanation for all forms of technology made or used in living systems.

The proposed new concept here differs from current approaches and seeks to explain technology as a system in interaction with living systems to solve problems and achieve specific goals. This new approach here can also facilitate the identification of a greater variety of forms of technology that may never have been considered, which could broaden the understanding of characteristics and behavior of technological innovation and technological advancement. To sum up, the suggested theory here has the potential to generate new theoretical and empirical predictions.

# A PROPOSED GENERAL DEFINITION OF TECHNOLOGY WITH A SYSTEMIC-PURPOSEFUL PERSPECTIVE

◻ *Philosophical foundations of technology with a systemic-purposeful perspective*

Although definitions of technology exist to explain the patterns of technological innovations (Sahal, 1981), there is no general definition that can explain the emergence and evolution of technology in a context of complex interaction between technology and human and other animal species. In order to define the concept of technology in this context, it is useful to explain complexity and complex systems (cf., Barton, 2014). Simon (1962, p. 468) states that: "a complex system [is]… one made up of a large number of parts that interact in a nonsimple way …. complexity frequently takes the form of hierarchy, and …. a hierarchic system … is composed of interrelated subsystems, each of the latter being, in turn, hierarchic in structure until we reach some lowest level of elementary subsystem." McNerney et al. (2011, p. 9008) argue that: "technology can be decomposed into $n$ components, each of which interacts with a cluster of $d-1$ other components." This modularity can be one of the most important features of technology as complex adaptive systems to describe the use of common units and to create product or process variants (cf., Arthur, 2009; Bryan et al., 2007; Huang and Kusiak, 1998; Mazzolini et al., 2018; Ulrich, 1995; Ulrich and Eppinger, 2012). Another characteristic of complex systems is the interaction between systems and sub-systems, such that the hierarchy can be defined in terms of the intensity of interaction between elements of the system. A distinction in hierarchic systems is the interaction between systems and the interaction within systems—i.e., among the parts of those systems (cf., AlGeddawy and ElMaraghy, 2013; Kashkoush and ElMaraghy, 2015). In this setting, Simon (1962, p. 474) points out that hierarchies have the property of nearly decomposable systems: "(a) in a nearly decomposable system, the short run behavior of each of the component subsystems is approximately independent of the short-run behavior of the other components; (b) in the long run, the behavior of any one of the components depends in only an aggregate way on the behavior of the other components."

Schuster (2016, p. 8) argues that: "Technologies form complex networks of mutual dependences just as the different species do in the food webs of ecosystems" (cf. also, Iacopini et al., 2018; Mazzolini et al., 2018; Vespignani, 2009).

Bunge (1990, pp. 231-232) argues that: "technology may be regarded as the field of knowledge concerned with designing artifacts and planning their realization, operation, adjustment, maintenance and monitoring in the light of scientific knowledge. (an artifact can be a thing, … or a process, and that it can be physical, chemical, biological, or social.)". Bunge (1990, p. 231, original emphasis) also claims that:

> A family of technologies is a system ***T*** every component of which is representable by an eleven-tuple ***T***= <C, S, D, G, F, B, P, K, A, M, V> ….C = a professional *Community* within, S = a larger *Society,* D = *Domain* of objects, natural, artificial, social, G = *General outlook* or philosophy: epistemologically realist but also pragmatic, F = Formal background of logic and mathematics, B = *specific Background of data, hypotheses, methods, and designs of* related fields, P = *Problems*, all related to D or some other item in the set, K = *Knowledge*: data, hypotheses, and designs of the field, A = *Aims*, especially inventing new artifacts or new uses for old (including social) artifacts, M = *Methods*, both scientific and technological, V = *Values*, especially the value of using science and technology for the benefit of society and (1) there is always at least one other partially overlapping family of technologies; and (2) the sets change over time as a result of their own R&D activities.

Bunge (1990) argues that this definition presupposes an approach that identifies systematization with an exact—namely mathematical—formulation in the manner of theorizing within pure science (cf., Coccia, 2018b; 2019b; Coccia and Wang, 2016). Moreover, Bunge (1990) states that general systems theory cannot alone solve any particular problem, but it can help pose problems—identifying their components, couplings among these components, and relations to an environment—in ways that make solutions more likely (Coccia, 2005, 2008). In this context, Bunge (1990) shows examples, including the general theory of machines, automata theories, cybernetics, etc.

In addition to systems approach, it is important to clarify the philosophical aspects of purposive behavior. Singer (1947) shows that teleological concepts are extremely fruitful in the study of machine behavior, and that such concepts can be also treated experimentally (cf., Churchman and Ackoff, 1950; Rosenblueth and Wiener, 1950; Rosenblueth et al., 1943). In this context, Ackoff (1971, p. 666) introduces the concept of a purposive system that is a multi-goal-seeking system with different goals having a common property: system's purpose. This type of system can pursue different goals but it does not select the goal to be pursued. The goal is determined by the initiating event and the system does choose the means by which to pursue its goals. In addition, Ackoff (1971) also introduces the concept of purposeful system that can produce the same outcome in different ways in the same (internal or external) state and can produce different outcomes in the same and different states. Thus a purposeful system can change its goals under constant conditions; i.e., it selects ends as well as means and thus displays will. Human beings are the most familiar examples of such systems. This philosophical background is essential for suggested definition of technology.

❑ *A proposed general definition of technology*

The primary goal of this study is to define the concept of technology; and that definition should meet the conditions of independence, generality, epistemological applicability and empirical correctness (Brandon, 1978). In philosophy of science, definitions can be of two types, descriptive and stipulative. (Hempel, 1966). Descriptive definitions simply describe the meaning of terms already in use; stipulative definitions assign, by stipulation, special meaning to a term. The study here endeavors to suggest a stipulative definition of technology with a perspective based on interaction between a technology and living systems that make and use technology for the purpose of adaptation in environment.

The *proposed definition of technology* here is that:

> Technology is a complex system of artifact, made and/or used by living systems, that is composed of more than one entity or sub-system and a relationship that holds between each entity and at least one other entity in the system, selected considering technical and economic characteristics, to satisfy needs, achieve goals and/or solve problems of users for the purposes of adaptation and/or survival in environment. Technology changes current modes of cognition and action to enable makers and/or users to take advantage of important opportunities or to cope with consequential environmental threats.

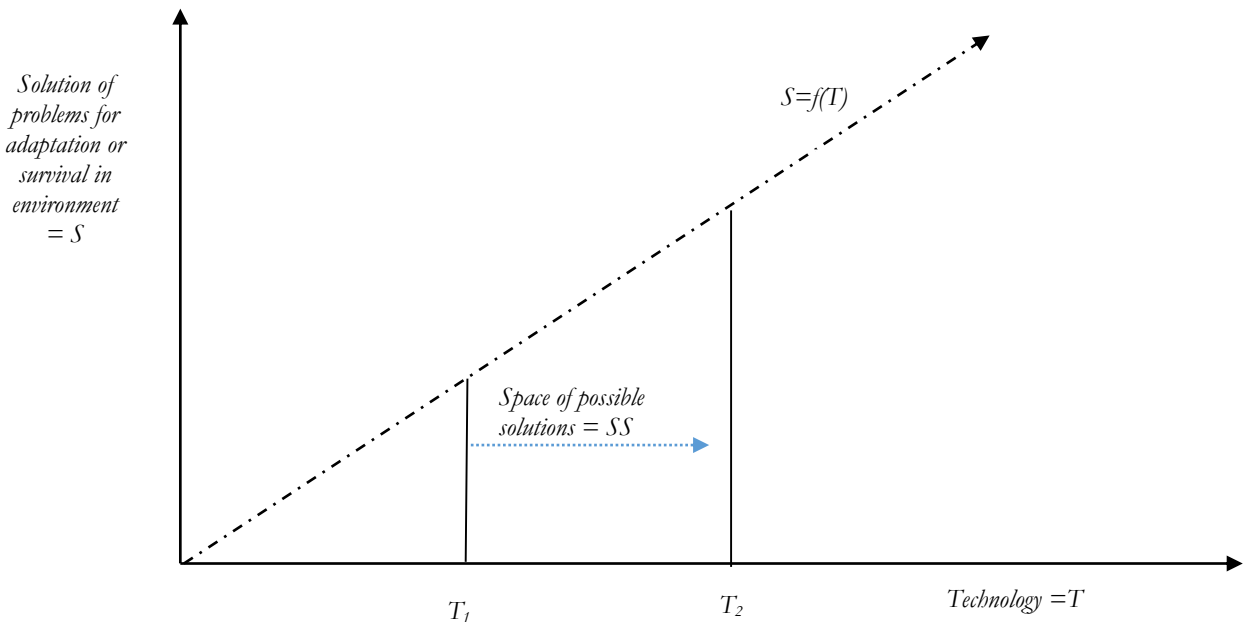

*Figure 1*. Technology extends the space of possible solutions of users for adapting and surviving in environment.

Overall the role of adaptation and survival of adopters can be a vital driver of technology creation and application[1] to take advantage of important opportunities or to cope with environmental threats, extending the space of possible solutions *SS (Fig. 1)*

$$SS = \int_{T_1}^{T_2} f(Tt)dT$$

*Remark*: technology as a complex system $T_1$, just defined, is formed by different elements given by incremental and radical innovations. Technological change is the progress of technology from a system $T_1$ to $T_2, T_3 \ldots$ with advances of new technological trajectories and technological paradigms to achieve specific goals and/or solve problems with effects in environment and society (cf., Coccia, 2015a, 2015b, 2016, 2018). In short, technological change is driven by clusters of radical and incremental innovations (Figure 2).

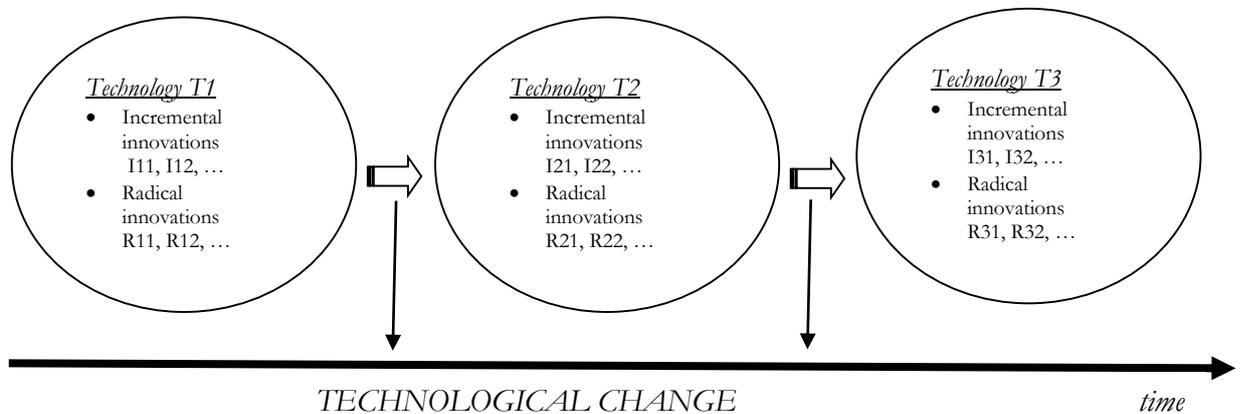

*Figure 2.* Technology, elements of technology and technological change

---

[1] For other socioeconomic determinants of technology see Coccia, 2012, 2014, 2015, 2017, 2017a, 2018, 2018c, 2018d, 2018e; Coccia and Wang, 2015.

*General properties of technology* are:

1. Property of not independence of any technological innovation: the long-run behavior and evolution of any technological innovation $Ti$ is not independent from the behavior and evolution of the other technological innovations $Tj$, $\forall i = 1, \ldots, n$ and $\forall j = 1, \ldots, m$ (cf., Coccia, 2018a, 2019, 2019a). In general, technologies do not function as independent systems themselves, but *de facto* they depend on other technologies and systems to form a complex system of inter-related parts that interact in a non-simple way (cf., Schuster, 2016, p. 8).

2. Property of maximization of mutual benefaction: selection processes, based on technical and economic criteria, during the interaction between technologies and human/animal species-technology reduce negative effects and favor positive effects directed to an evolution of reciprocal adaptations of technologies in environment to satisfy needs and solve problems[2].

3. Coevolution of technologies is the evolution of reciprocal adaptations in a complex system, supporting the reciprocal enhancement of technologies' growth rate and innovation—i.e., a modification and/or improvement of technologies based on interaction and adaptation in complex systems to satisfy changing needs and solve consequential problems in environment.

☐ *Example of technology in human society and in other animal species*
In agriculture, the plowing is one of the most energy-consuming operations (Walker, 1929). The farm tractor, produced and used by human being, is a complex system as defined above. This technology, because of technical characteristics, has been selected and adopted in agricultural environment, generating a substitution of mechanical for animal power to satisfy needs of reducing energy-consumption operations for humans, supporting a higher productivity in agricultural production for human and animal nutrition. In fact, farm tractor is a general-purpose technology in agriculture to take advantage of important opportunities in plowing and a wider range of farm operations (Sahal, 1981). Moreover, farm tractor is a technological system formed by various major innovations, such as pneumatic tires, quality of fuels, and minor innovations, such as durable valve, piston, etc. (Sahal, 1981). A technical change in farm tractor is from gasoline track-type tractors to Diesel-powered track type tractor.

---

[2] May (1981, p. 95) suggests the concept of "orgy of mutual benefaction" that may be also appropriate for explaining the interaction within technological domains.

For animal species there are some examples of technology, such as beaver dams; McGrew (2013) argues that the New Caledonian crow has been championed as being equivalent or superior to the ape in elementary food-getting technology. These example of elementary technologies in animal species can be embodied in suggested definition above.

❐ *Requirements of the definition of technology for the philosophy of science*

1. Independence

    The suggested definition of technology explains the structure and goal of technology in interaction with living systems. If the relational concept of technology here has to play its explanatory role in studies of technology, it should not become a tautology. This requirement is called condition of independence.

2. Generality

    The proposed definition seems to be general, i.e., universally applicable throughout all material and not material artifacts, systemic and not systemic

technologies, etc. both for humans and other animal species.

3. Epistemological applicability

   The suggested definition is not vague, such that it can be applied to particular cases and it is testable.

   In order to satisfy this requirement, the second goal of this study is to operationalize the concept of technology for practical purposes.

   Suppose the simplest possible case of only two technologies, *H* and *P*, forming a complex and purposeful system *T*(H, P); of course, the model can be generalized for complex systems including many subsystems of technology.

a) Let *P(t)* be the extent of technological advances of a technology *P* at the time *t* and *H(t)* be the extent of technological advances of a technology *H* that interacts with *P* at the same time in a complex system.

b) Suppose that both *P* and *H* evolve according to some *S*-shaped pattern of technological growth, such a pattern can be represented analytically in terms of the differential equation of logistic function.

   For *H*, the starting equation is:

   $$\frac{1}{H}\frac{dH}{dt} = \frac{b_1}{K_1}(K_1 - H) \qquad [1]$$

   *Mutatis mutandis*, for technology *P(t)* the equation is:

   $$\log\frac{K_2 - P}{P} = a_2 - b_2 t \qquad [2]$$

   The logistic curve here is a symmetrical *S*-shaped curve with a point of inflection at 0.5K with $a_{1,2}$ are constants depending on initial conditions, $K_{1,2}$ are equilibrium levels of growth, and $b_{1,2}$ are rate-of-growth parameters (1= technological system *H*, 2= technological subsystem *P*).

Solving equations [1] and [2] for *t*, the result is:

$$t = \frac{a_1}{b_1} - \frac{1}{b_1}\log\frac{K_1 - H}{H} = \frac{a_2}{b_2} - \frac{1}{b_2}\log\frac{K_2 - P}{P}$$

The expression generated is:

$$\frac{H}{K_1 - H} = C_1\left(\frac{P}{K_2 - P}\right)^{\frac{b_1}{b_2}} \qquad [3]$$

The concept of technology as system *H* in interaction with a subsystem *P* directed to achieve goals and solve problems, it can be represented with following equation (cf., Coccia, 2019a), given by:

$$P = A\,(H)^B \qquad [4]$$

The logarithmic form of the equation [4] is a simple linear relationship:

$$\log P = \log A + B\,\log H \qquad [5]$$

*B* is the evolutionary coefficient of growth that measures the evolution of technological subsystem *P* in relation to technological system *H* to achieve specific goals fixed by living systems.

To apply this model, based on systemic-purposeful perspective of technology, it is important to consider Functional Measures of Technology (FMT) that are the technical characteristics of technologies and their change can indicate the evolution of technology over the course of time based on major and minor innovations, such the measure of fuel-consumption efficiency of vehicles (cf., Sahal, 1981, pp. 27-29).

A practical example is electricity generated by internal-combustion plants; FMTs of this technology over 1920-1970 period in US market are:

1. Average fuel-consumption efficiency in kilowatt-hours per cubic foot of gas indicates the technological advances of boiler, turbines and electrical generator (a subsystem of internal combustion plant). This FMT represents the dependent variable *P* in the model [5].

2. Average scale of plant utilization (the ratio of net production of electrical energy by internal-combustion type plants in millions of kilowatt-hours to total number of these plants) indicates a proxy of technological advances of plants with internal-combustion technology. This FMT represents the explanatory variable of the technology *H* in the model [5].

Table 1 – Estimated relationship for internal-combustion plants with gas turbines (1920-1970 period in US market)

| *Dependent variable*: *log* Average fuel consumption efficiency in kwh per cubic feet of gas (*P*=technological advances of turbine and various equipment) | | | | |
|---|---|---|---|---|
| | *Constant* $\alpha$ (St. Err.) | *Evolutionary coefficient* $\beta=B$ (St. Err.) | $R^2$ *adj.* (St. Err. of the Estimate) | *F* (sign.) |
| Gas turbine and various equipment | −2.93*** (0.02) | 0.35*** (0.02) | 0.81 (0.14) | 213.63 (0.001) |

*Note*: ***Coefficient β is significant at 1‰; Explanatory variable is *log* Average scale of internal-combustion plants (Host technology *H*)

Table 1 shows estimated relationship with Ordinary Least Squares method of electricity generation with internal-combustion plants having gas turbines; the coefficient of evolutionary growth of this technology is B = 0.35, i.e., B < 1. In short, the technology in the generation of electricity in internal-combustion plants, as complex system, evolves with a low evolutionary pathway of underdevelopment over the course of time (cf., Coccia, 2019a).

4. Empirical correctness

The proposed definition of technology may be empirically correct, i.e. to fit the fact of artifact and techniques in environment. The suggested definition seems not be *false* or more precisely seems to be nontautologously *true*. For instance, insulin that the pancreas produces for metabolism of the body is not a technology, whereas insulin as drug for human versions can be made either by modifying pig versions or recombinant technology, such as transgenic plants are very attractive expression system, which can be exploited to produce insulin as technological drug in large quantities for therapeutic use in human societies (cf., Baeshen et al., 2014).

**DISCUSSION AND CONCLUSIVE OBSERVATIONS**

The concept technology has been one of the most troublesome and yet one of the most important concepts in science. Defining concepts in science is a vital scientific activity because a scientific field can develop only on the base of new comprehensive concepts. Scientists should open the debate regarding the nature of technology based on interaction between technology and living systems that may explain and generalize vital aspects of technology, evolution of technology and technical change for adaptation of users in changing environment (cf., Pistorius and Utterback, 1997; Utterback et al., 2019).

The study here proposes the definition of technology in a theoretical framework of systems and purposive behavior. On the basis of theoretical and empirical analysis presented in this study, proposed definition of technology seems to clarify and generalize, whenever possible, some universal characteristics of technology. In particular, the results of scientific analyses here reveal that:

1. Long-run behavior of any technology is *not* independent of the living systems (human and other animal species) as well as of other inter-related technologies.
2. Technologies, during the interaction with living systems and other technologies, reduce negative effects and favor positive effects in the long run directed to an evolution of reciprocal adaptations of technologies in environment.

4. Technologies co-evolve with the evolution of reciprocal adaptations in a complex system, supporting the reciprocal enhancement of technologies' growth rate and innovation in environment.

The study documented here makes a unique contribution, for the first time to our knowledge, by suggesting a general definition of technology useful for natural and social sciences. In this context, humans act as ecosystem engineers able to change the socioeconomic environment and support progress (cf., Solé et al., 2013). The definition of technology presented in the study here is adequate in some cases but less in others because of the diversity of technologies and their interaction with users and ecosystems (cf., Coccia, 2018; Pistorius and Utterback, 1997). In fact, a definition of technology that satisfies all four desiderata (independence, generality, epistemological applicability and empirical correctness, cf., Brandon, 1978) is a difficult task because of a trade-off between desiderata, such as between testability and systematic unification of a definition. Nevertheless, the definition here seems to keep its validity in explaining several phenomena of the origin and evolution of technology for supporting the adaptation and survival of living systems in normal and aversive environment. New definition of technology suggests some general properties that are a reasonable starting point for understanding the universal features of technologies that lead to technological and economic change, though we know, *de facto*, that other things are often not equal over time and space in the domain of technology.

Overall, then, the proposed definition of technology may lay the foundation for development of more sophisticated concepts and theoretical frameworks as well as to encourage further theoretical exploration in the *terra incognita* of the interaction among technologies and living systems to generalize further properties of the nature and evolution of technology. To conclude, the concept of technology is still being revised and debated and indicates that we have some way to go before we can say that we know why animals use tools, and why humans became so dependent on them. To resolve this scientific problem, we need input from the varied scientific fields. We also need high-quality research data from many more technologies and tool-using species: studies that aim to identify commonalities and differences between technologies and instruments because of ecological drivers, cognitive or morphological factors, or factors of social learning. Future efforts in this research field will be also directed to provide further empirical evidence, also considering dependency-network framework to better evaluate this new definition with other properties about behavior of technology and technological evolution for adopters in complex environment. Hence, identifying generalizable definition of technology at the intersection of engineering, economics, psychology. sociology, anthropology, and perhaps ethology and human biology is a non-trivial exercise. In fact, Wright (1997, p. 1562) properly claims that: "In the world of technological change, bounded rationality is the rule."